# Prediction of Discharge Capacity of Labyrinth Weir with Gene Expression Programming

Hossein Bonakdari [1*], Isa Ebtehaj [2], Bahram Gharabaghi [3], Ali Sharifi [4], Amir Mosavi [5]

[1]Department of Soils and Agri-Food Engineering, Laval University, Québec, G1V0A6, Canada
[2]Department of Civil Engineering, Razi University, Kermanshah, Iran
[3]School of Engineering, University of Guelph, Guelph, Ontario, NIG 2W1, Canada
[4]Department of Statistics, Razi University, Kermanshah, Iran
[5] Department of Mathematics and Informatics, J. Selye University, 94501 Komarno, Slovakia
*Corresponding author, Phone: +1 418 656-2131, Fax: +1 418 656-3723, E-mail: hossein.bonakdari@fsaa.ulaval.ca

**Abstract.** This paper proposes a model based on gene expression programming for predicting discharge coefficient of triangular labyrinth weirs. The parameters influencing discharge coefficient prediction were first examined and presented as crest height ratio to the head over the crest of the weir (p/y), crest length of water to channel width (L/W), crest length of water to the head over the crest of the weir (L/y), Froude number (F=V/√(gy)) and vertex angle ($\theta$) dimensionless parameters. Different models were then presented using sensitivity analysis in order to examine each of the dimensionless parameters presented in this study. In addition, an equation was presented through the use of nonlinear regression (NLR) for the purpose of comparison with GEP. The results of the studies conducted by using different statistical indexes indicated that GEP is more capable than NLR. This is to the extent that GEP predicts discharge coefficient with an average relative error of approximately 2.5% in such manner that the predicted values have less than 5% relative error in the worst model.

## 1    Introduction

Conventional weirs are structures used to control, regulate and measure water level and flow volume in irrigation and drainage networks and water and wastewater treatment plants. A conventional weir is usually installed along the flow and perpendicular to channel axis. Conventional weirs include rectangular, V-notch, labyrinth and complex weirs. Many theoretical and experimental studies investigated passing flow from conventional weirs. Taylor [1] presented an experimental study on hydraulic labyrinth weirs. Hay and Taylor [2] described how the head on the labyrinth weir effects the discharge ratio. Tullis et al. [3] investigated trapezoid labyrinth weirs and indicated that their discharge capacity was a function of total head, effective length of weir crest and coefficient of discharge of labyrinth weir. Wormleaton and Soufiani [4] studied hydraulic features and aeration of triangle labyrinth weirs. They found that aeration efficiency of triangle labyrinth weirs is more than linear weirs with equal length. Also, Wormleaton and Tsang [5] studied aeration of rectangular weirs experimentally. Emiroglu and Baylar [6] investigated the effects of weir included angle and water sill slope of weir on aeration in triangle labyrinth weirs. Tullis et al. [7] studied hydraulic behavior and flow head on submerged labyrinth weirs. They concluded that the flow over submerged labyrinth weirs did not depend on labyrinth weir sidewall angles. Bagheri and Heidarpour [8] used free vortex theory to estimate discharge coefficient of sharp-crested rectangular weirs as a function of flow features, channel geometry and conventional weir. Kumar et al. [9] experimentally investigated discharging capacity of triangle labyrinth weirs. They suggested a relation to calculate the flow over triangle labyrinth weirs through analyzing experimental data.

Considering the complexity of engineering problems and the growing number of engineering studies, new methods called soft computing, were significantly used during recent decade that were more efficient and more accurate in solving complicated and difficult engineering issues and, facilitating studies [10-13]. Soft computing and artificial intelligence were used by different researchers to estimate and predict different hydraulic and hydrologic problems especially discharge coefficient [14-17]. Emiroglu et al. [18] used Adaptive Neuro Fuzzy Inference System (ANFIS) techniques to predict discharge capacity of the triangular labyrinth side weir. They introduced an equation for discharge coefficient in this type of side weirs. The diversion flow passing over sharp-crested rectangular side weirs were predicted using Feed Forward Neural Networks (FFNN) and Radial Basis Neural Networks (RBNN) by [19]. Bilhan et al. [19] introduced an equation for discharge coefficient as a function of geometric and hydraulic features for sharp-crested rectangular side weirs. Emiroglu et al. [20] used artificial neural networks to introduce a relation which calculated discharge coefficient of triangle labyrinth weirs located in rectangular in under critical flow conditions.

Gene Expression Programming (GEP) is one method used in water hydraulic engineering during recent years. Unlike artificial neural system and neuro fuzzy systems which include a black box, the suggested method showed high accuracy in estimating the given parameter and relation [21-25].

Using Gene Expression Programming (GEP), the present study aims to introduce an equation to predict discharge coefficient. Therefore, the parameters influencing discharge coefficient are first determined and then an equation is presented using GEP. Following that, the effect of each of the dimensionless parameters is examined on predicting discharge coefficient through using sensitivity analysis. Also, the results of the GEP model are compared with that of nonlinear regression (NLR).

## 2    Data collection

The present study used Kumar et al. [9] experimental data to estimate the coefficient of discharge. A horizontal rectangular channel with 12 m length, 0.28 m width and 0.41 m depth was used in their tests. The used triangle weir was located 11 m away from the channel entrance. Water was provided for the channel through an inlet pipe from an overhead tank supplied with an overflow arrangement to keep a constant head. The water height over weir crest was measured by point gages having ±0.1 mm accuracy. Ventilation holes were installed on both sides of the weir's downstream for the purpose of aeration of the nappe. Wave suppressors and Grid walls were structured at the upstream of the channel to break and dissipate the surface disturbances and to enlarge the size of eddies, respectively. They conducted their experiments on 30, 60, 90, 120, 150, and 180 degree weirs. They also used varied discharges for each of the mentioned angles. They eventually carried out 123 different experiments for different discharges and angles. Schematic of Kumar et al. [9] experimental model is illustrated in Fig. 1. Table 1 shows the parameters used in the present study.

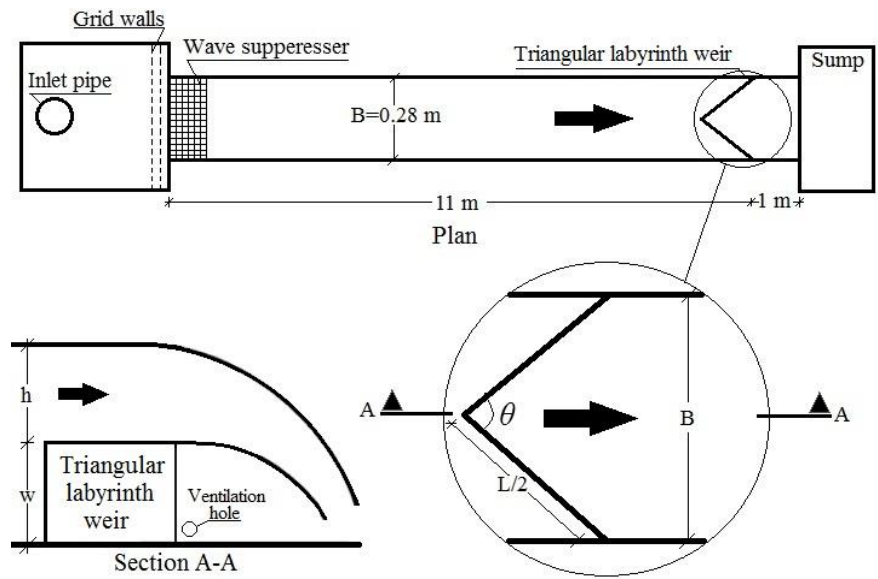

**Fig. 1.** Schematic of Kumar et al. [9] experimental model

Table 1. Parameters used to estimate discharge coefficient [9]

|     | p/y   | L/W   | F     | W/y   | θ (degree) | $C_d$ |
|-----|-------|-------|-------|-------|------------|-------|
| min | 0.581 | 1     | 0.608 | 1.62  | 30         | 0.54  |
| max | 0.92  | 3.864 | 3.261 | 10.82 | 180        | 0.906 |

## 3  Overview of Gene Expression Programming

GEP is a developed genetic programming [26]. It is a search technique relying on computer programs such as decision tree, logical expressions, polynomial construct, and mathematics statements. GEP computer programs are coded as line chromosomes and the final presentation is in the form of expression trees (ETs) [27]. ETs are complex computer programs which are developed to solve a given problem and are selected according to their fitness to the problem [25]. Considering that in GP, genotype and phenotype are mixed in a simple replicator system, GEP of a genotype/phenotype system is developed where genotype is completely separated from phenotype. Therefore, developed GEP genotype/phenotype system is 100 to 60000 times more effective than GP system [28, 29].

In GEP process, the first chromosome of each independent parameter is randomly generated in the population. Then, they are developed and all independent parameters are evaluated based on fitness function and are used as a part to produce new generation with different characteristics. People of the new generation develop through confrontation with the selection environment, expression of the genomes and reproduction with modification. The process continues until getting the predefined generation or getting the answer [28, 29].

Ferreira [30] described the fitness of an individual function (i) for the fitness model (j) as: $If\ \ E(ij)\ \leq p, then\ f_{(ij)} = 1,\ ,else\ \ f_{(ij)} = 0$
(1)

where p and E(ij) are the precision and error, respectively. Then the absolute error can be obtained from:

$$E(ij) = |p_{(ij)} - T_j| \qquad (2)$$

Where the ($f_i$) for an individual function calculated as follows:

$$f_i = \sum(R - |p_{(ij)} - T_j|) \qquad (3)$$

where $T_j$, R and $p_{(ij)}$ are the target values, selection range, and predicted values, respectively. Accordingly, the terminal set (T) and function set (F) are calculated to select the chromosomes. Fig. 2 presents the GEP flowchart.

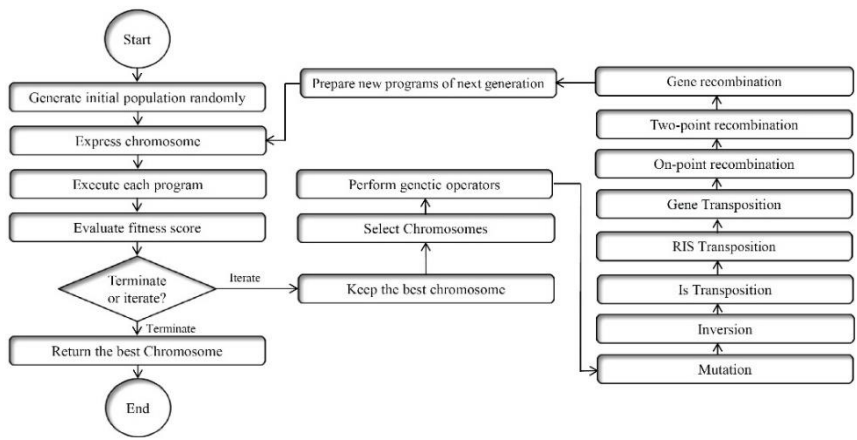

**Fig. 2**. Gene Expression Programming flowchart

## 4    Derivation discharge coefficient based on GEP

Reviewing the recent studies conducted on estimating discharge coefficient in weirs, crest height (p), head over the crest of the weir (y), crest length of the weir (L), channel width (W), and Froude number (F=V/√(gy)) parameters can be named [19,18,20, 31]. The dimensionless parameters in estimating discharge coefficient can be presented as equation 4 through using dimensional analysis.

$$C_d = f(\frac{w}{y}, \frac{L}{b}, \frac{L}{y}, F, \theta) \tag{4}$$

The manner of function estimation through using the GEP method to predict discharge coefficient will be presented in this section. For training 20% of data set is used randomly as suggested by Kumar et al. [9]. Furthermore, 80% of data can be used for testing. To produce an initial population of, according to Ferreira's [28] the range of 30-100 is suggested In the next step a fitness function is calculated using MSE as follows:

$$f_i = \frac{100}{1+E_i} \quad \text{for} \quad E_i = p_{ij} - O_j \tag{5}$$

where $P_{ij}$, and $Q_{ij}$ represent the predicted and fitness case values for i individual chromosome for fitness case j. The set of terminals are developed as follows:

$$T = \{C_d, \frac{w}{y}, \frac{L}{b}, \frac{L}{y}, F, \theta\} \tag{6}$$

Where the number of genes and their head and tail length are calculated for every chromosome. In the present study, three genes were used in each chromosome. In this study, the {+} operator is utilized to link function among the genes. The {x} function presented in Table 2 provides the (1-x) amount. Using equation (4) and the expression tree presented in Fig. 3, the model presented by using GEP can be expressed as equation (7); its parameters' values are presented in Table 3.

$$C_d = Exp\left[F - \frac{L}{b} + 1.8\right] - Exp\left[1 - Exp\left[\frac{w}{y}\right]\right] + \frac{w}{y} \times Exp\left[0.034\frac{L}{y}(\theta - 1)\right] +$$

$$1 - \left[\frac{w}{y} + Exp\left[\frac{L}{b} + 1.58F - \theta + 1.79\right]\right]$$

$$\tag{7}$$

where $C_d$ is coefficient of discharge, w/y the ratio of crest height to head over the crest of the weir, L/W ratio of crest length of water to channel width, L/y the ratio of crest length of water to the head over the crest of the weir, F, Froude number and $\theta$ vortex angle.

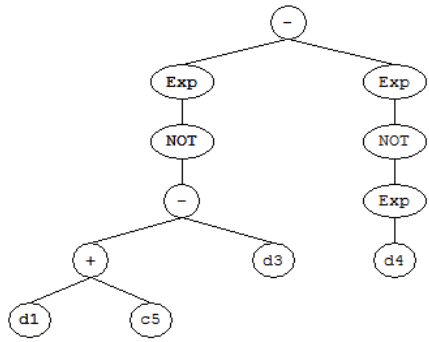

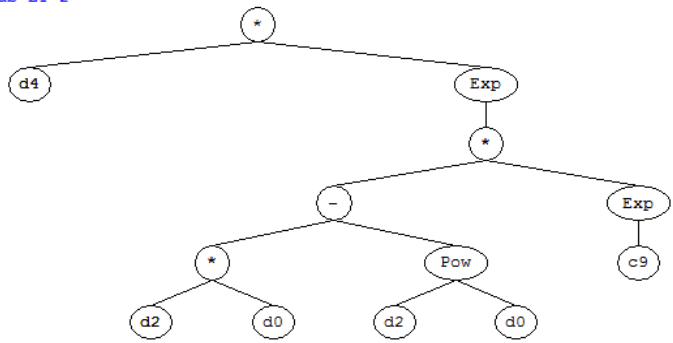

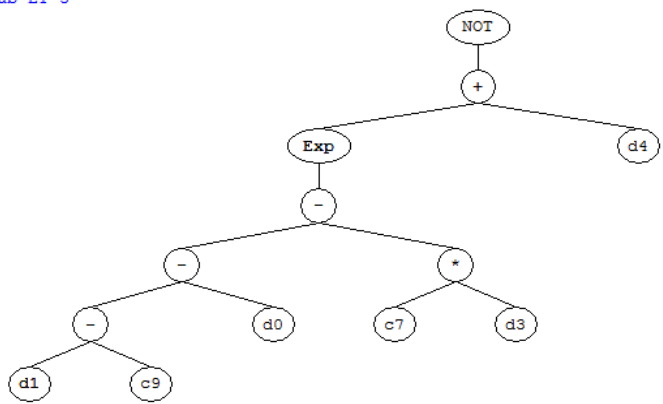

**Fig. 3.** Expression tree (ET) for presented model (Equation 7)

**Table 2.** Parameters of GEP model

| Parameter | Setting |
|---|---|
| Population size | 50 |
| Number of generations | 40000 |
| Number of chromosomes | 40 |
| Number of genes | 3 |
| function set | ×, -, +, Not, Exp, Pow |
| Linking function | Addition |
| Mutation rate | 0.0014 |
| Inversion rate | 0.05 |
| IS transposition rate | 0.15 |
| RIS transposition rate | 0.15 |
| Gene transposition rate | 0.20 |
| One-point recombination rate | 0.15 |
| Two-point recombination rate | 0.15 |
| Gene recombination rate | 0.30 |

**Table 3.** The values of the parameters used in ET (Fig. 3)

| Parameter | value | Parameter | value |
|---|---|---|---|
| d0 | θ | G1C5 | 2.8 |
| d1 | L/W | G2C9 | -3.38 |
| d2 | L/y | G3C7 | 1.58 |
| d3 | F | G3C9 | 1.79 |
| d4 | p/y | - | - |

## 5    Result and discussion

The accuracy of the model presented through the use of GEP (equation 7) is examined in this section with using different statistical indexes. In addition, sensitivity analysis is also conducted in order to study the effect of each of the dimensionless parameter presented in predicting discharge coefficient. Following that, the results from this model will also be compared with the results of the nonlinear regression analysis (NLR) to examine the accuracy of the model presented by using GEP.

In order to verify the accuracy of the estimated model at each step of model development, the results of analysis of GEP and NLR is based on the criteria of the coefficient of determination ($R^2$), Root Mean Square Error (RMSE), Mean Absolute Percentage Error (MAPE), Adjusted Coefficient of Efficiency (CE) and Scatter Index (SI) as defined in the following forms:

$$R^2 = \left[ \frac{\sum_{i=1}^{n}\left(C_{d_{EXP_i}} - \overline{C_{d_{EXP}}}\right)\left(C_{d_{GEP_i}} - \overline{C_{d_{GEP}}}\right)}{\sqrt{\sum_{i=1}^{n}\left(C_{d_{EXP_i}} - \overline{C_{d_{EXP}}}\right)^2 \sum_{i=1}^{n}\left(C_{d_{GEP_i}} - \overline{C_{d_{GEP}}}\right)^2}} \right]^2 \tag{8}$$

$$RMSE = \sqrt{\left(\frac{1}{n}\right)\sum_{i=1}^{n}\left(C_{d_{EXP_i}} - C_{d_{GEP_i}}\right)^2} \tag{9}$$

$$MAPE = \left(\frac{1}{n}\right)\sum_{i=1}^{n}\left(\frac{\left|C_{d_{EXP_i}} - C_{d_{GEP_i}}\right|}{C_{d_{EXP_i}}}\right) \times 100 \tag{10}$$

$$CE = 1 - \frac{\sum_{i=1}^{n}\left|C_{d_{EXP_i}} - C_{d_{EXP_i}}\right|}{\sum_{i=1}^{n}\left|C_{d_{EXP_i}} - \overline{C_{d_{EXP}}}\right|} \tag{11}$$

$$SI = \frac{RMSE}{\overline{C_{d_{EXP}}}} \tag{12}$$

where $C_{d_{EXP_i}}$ and $C_{d_{GEP_i}}$ denote the actual and modeled discharge coefficient values and $\overline{C_{d_{EXP}}}$ and $\overline{C_{d_{GEP}}}$ represent the mean actual and modeled discharge coefficient values, respectively.

The closer the value of index $R^2$ to 1, the more it shows the compatibility of the estimated value with the real value. Results which are obtained from coefficient of determination ($R^2$) have been simulated in relation with linear dependence between real and corresponding values (for the present case, the actual and simulated discharge coefficient values) and they are sensitive towards deviated points; so in evaluating the results, we cannot solely rely on this index. Thus, other statistical indexes like mean absolute percentage error (MAPE) - which shows the difference between real and estimated models in form of percentage of actual values- and root mean square error (RMSE) - which considers the weight of larger errors by powering the difference between actual and estimated values - are needed in order to estimate the function of the models. Both MAPE and RMSE indexes can include zero value (best mode) and infinity (worst value). Also, dimensionless RMSE criterion which is stated in SI form can be applied in estimating different models without considering dimension of parameters. Besides, as a complementary criterion, the "adjusted coefficient efficiency (CE)" could be utilized for evaluating the precision of models. This index reports the difference between the proportion of remainders variance (numerator term) and the data variance (denominator term) from 1. If this index equals 1, the presented model has done data estimation in the best way. Simultaneous use of these indexes could provide sufficient information for precision of the applied models [32].

As mentioned earlier, the data utilized in this study is divided into two groups of "train" and "test" in such way that 20% of the data is selected through random selection without replacement for the purpose of testing, and the discharge coefficient parameter was presented as equation (7) using the remaining 80% data. Fig. 4 shows the results obtained from training the presented GEP model in test and train states. The x axis indicates the actual values and y axis presents the values predicted by GEP. It could be seen in the figure that almost the majority of the predicted amounts predict the discharge coefficient fairly accurately in both states of test and train. The GEP model presented in the train predicts the train-state values with $R^2=0.95$ and an average relative error percentage approximate to 2% (MAPE). Most of the values presented in this state have a less- than- 5% relative error. The other statistical indexes used in the train state of this research are RMSE=0.017, CE=0.78 and SI=0.02 indexes MAPE and RMSE have very low amounts - as can be seen almost zero - which indicates the high accuracy of the presented model. The predicted values have an $R^2=0.93$ and a MAPE=2.53% in the test state which are almost similar to that of the train state. Also SI, CE, RMSE indexes are equal to 0.021, 0.67 and 0.029 respectively for the test state of this model. Therefore, considering Fig. 3 and the presented statistical indexes for train and test states of the presented GEP model, it could be stated that GEP predicts the discharge coefficient of triangular labyrinth weirs very well.

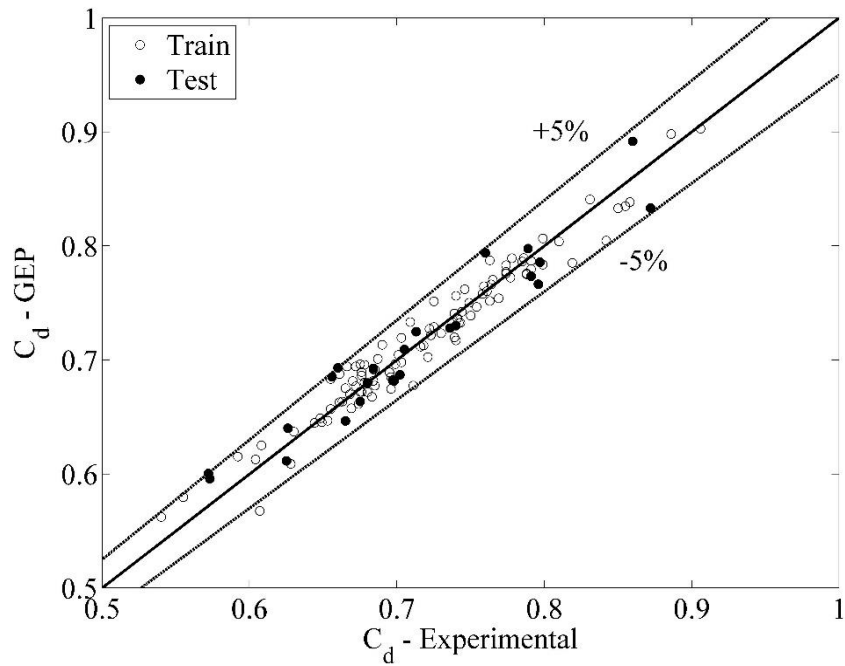

**Fig. 4.** Comparing estimated discharge coefficient with experimental result (Test and Train)

Through the use of sensitivity analysis in this section, the effect of each of the presented parameters is examined on predicting discharge coefficient of triangular labyrinth weirs. Therefore, different models are presented as Table 4. To estimate discharge in each of these models, the data is divided into two 80% and 20% groups, like they were in equation (7), for the purpose of training and testing the model, respectively. Tables 5 and 6 present the results of different statistical indexes, presented in the study, for the two "train" and "test" states, respectively. They demonstrate that the results of all the statistical indexes are better for model 1 when compared to the rest of the models for both train and test states. Also, Fig. 5 indicates that the maximum relative error of model 1is lesser than all other models. Therefore, it could be stated that the simultaneous use of dimensionless parameters of crest height ratio to the head over the crest of the weir (p/y), crest length of water to channel width (L/W), crest length of water to the head over the crest of the weir (L/y), Froude number (F=V/√(gy)) and vertex angle ($\theta$) is fixed in predicting discharge coefficient of rectangular labyrinth weirs. To examine the effect of each of the dimensionless parameters, the results of the statistical indexes of each model must be compared with regard to model 1 which is the best model and is presented as equation (7). It could be observed that model 2, which considers all the parameters of model 1 except for the vertex angle (θ), presents better results in comparison with models 3, 4, 5, and 6. Therefore, it could be stated that among the five presented dimensionless parameters, vertex angle (θ) parameter has the least value of effect on predicting discharge coefficient of triangular labyrinth weirs. Models 3, 4, 5 and 6 which disregard Froude number (F=V/√(gy)), crest length of water to the head over the crest of weir (L/Y), crest length of water to channel width (L/w), and crest height ratio to the head over the crest (p/y) dimensionless parameters respectively, do not present better results in comparison with models 1 and 2. Therefore, not using these parameters prevents predicting discharge coefficient relatively accurately in such manner that in some cases their maximum relative error is approximately 20% regarding Fig. 5. Therefore, it is essential to use these parameters in predicting discharge coefficient.

**Table 4.** Dependent parameters in discharge coefficient prediction

| Independent parameter | Dependent parameter | Model No. |
|---|---|---|
| p/y, L/W, L/y, F, $\theta$ | $C_d$ | 1 |
| p/y, L/W, L/y, F | $C_d$ | 2 |
| p/y, L/W, L/y, $\theta$ | $C_d$ | 3 |
| p/y, L/W, F, $\theta$ | $C_d$ | 4 |
| p/y, L/y, F, $\theta$ | $C_d$ | 5 |
| L/W, L/y, F, $\theta$ | $C_d$ | 6 |

**Table 5.** Statistics Indexes (Train)

|  | Model 1 | Model 2 | Model 3 | Model 4 | Model 5 | Model 6 |
|---|---|---|---|---|---|---|
| $R^2$ | 0.95 | 0.91 | 0.68 | 0.7 | 0.84 | 0.68 |
| RMSE | 0.017 | 0.021 | 0.055 | 0.040 | 0.028 | 0.039 |
| MAPE (%) | 1.920 | 2.442 | 6.139 | 4.379 | 2.823 | 4.452 |
| CE | 0.780 | 0.663 | 0.314 | 0.480 | 0.640 | 0.234 |
| SI | 0.020 | 0.029 | 0.076 | 0.055 | 0.039 | 0.054 |

**Table 6.** Statistics Indexes (Test)

|         | Model 1 | Model 2 | Model 3 | Model 4 | Model 5 | Model 6 |
|---------|---------|---------|---------|---------|---------|---------|
| $R^2$   | 0.93    | 0.88    | 0.73    | 0.76    | 0.88    | 0.63    |
| RMSE    | 0.021   | 0.026   | 0.054   | 0.040   | 0.028   | 0.047   |
| MAPE (%)| 2.538   | 3.004   | 6.142   | 4.891   | 3.056   | 5.327   |
| CE      | 0.699   | 0.652   | 0.375   | 0.505   | 0.665   | 0.202   |
| SI      | 0.029   | 0.037   | 0.076   | 0.055   | 0.039   | 0.065   |

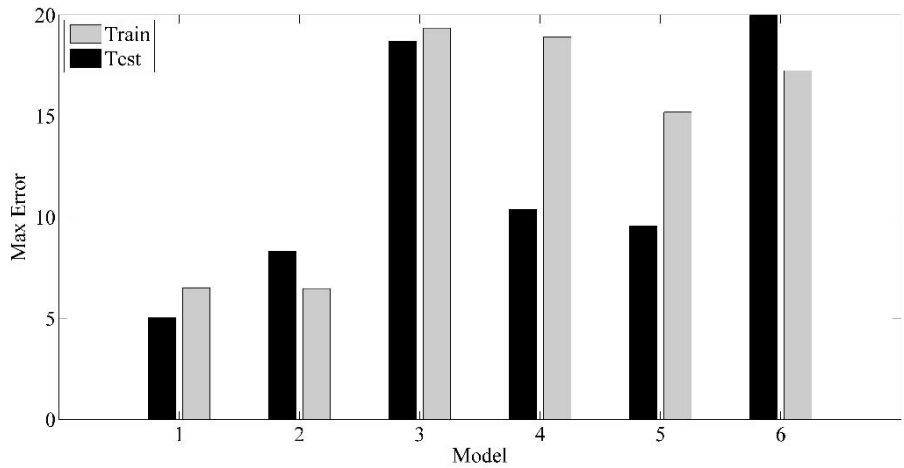

**Fig. 5.** Highest errors in six different models

Also, this study presents an equation (Eq. 13) that employs nonlinear regression (NLR) in MINITAB to predict discharge coefficient of triangular labyrinth. The set of data selected to train GEP were also used in this state in predicting the following equation. Also, through employing the data used by random selection without replacement for testing GEP, the accuracy of the following equation is used in this section.

$$C_d = 0.466 + 0.338 \left(\frac{p}{y}\right) - 0.183 \left(\frac{L}{W}\right) - 0.022 \left(\frac{L}{y}\right) + 0.31F + 0.12 sin(\theta)$$

(13)

Fig. 6 shows the results of discharge coefficient prediction for the two presented models using GEP and NLR. The x axis of this figure shows the experimental values (Target) and the y axis shows the values predicted through using GEP and NLR methods. The data used in this figure had no role in estimating equation (7) and (13) and as mentioned in the previous sections they were selected using random selection without replacement for the purpose of testing the model. The figure indicates that the equation presented by using GEP (equation 7) is fairly accurate in predicting discharge coefficient in a way that it predicts all the predicted discharge coefficients with a relative error less than 5%. This figure also shows that the equation presented by using

NLR mostly presents the discharge coefficient to be less than the actual value which leads to underestimating the prediction of the passing discharge and so causes underestimating. It could also be observed that the predicted values have a relative error greater than 5% in this state as opposed to GEP equation.

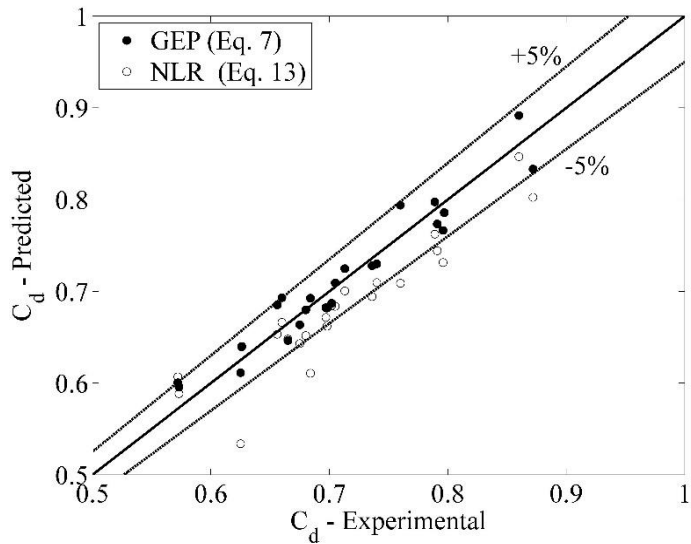

**Fig. 6.** Comparison of GEP and NLR in prediction of discharge coefficient of triangular labyrinth weirs (test)

Table 7 shows the results of the statistical indexes presented in this study in order to verify the accuracy of the equations presented by using GEP and NLR in predicting discharge coefficient for both states of train and test. Careful consideration of the table indicates that $R^2$ is more and less than 0.9 in both states of train and test of GEP and NLR respectively. It could also be seen that the average relative error is approximately 2.5% for GEP in test state and it is almost 4.5% for NLR. It is also observed that the results of RMSE and SI indexes for GEP are less than NLR and considering the fact that approaching these two indexes to zero indicates the higher accuracy of the model, it could be stated that the GEP model presented in this study is relatively less accurate with regard to the results obtained from NLR. The values predicted using equations (7), (GEP), and (13), (NLR), are presented in Table 8 for different hydraulic conditions.

**Table 7.** Comparing different statistical indexes for the discharge coefficients predicted by using GEP and NLR

| Statistics | Train | | Test | |
|---|---|---|---|---|
| Indexes | GEP (Eq.7) | NLR (Eq. 13) | GEP (Eq.7) | NLR (Eq. 13) |
| $R^2$ | 0.95 | 0.78 | 0.93 | 0.86 |
| RMSE  | 0.015 | 0.044 | 0.021 | 0.040 |
| MAPE (%) | 1.620 | 4.664 | 2.538 | 4.583 |
| CE    | 0.780 | 0.341 | 0.699 | 0.495 |
| SI    | 0.020 | 0.061 | 0.029 | 0.055 |

**Table 8.** Predicted coefficient of discharge using GEP and NLR

| θ (degree) | L (m) | w (m) | y (m) | Q (m³/s) | $C_d$ (Exp) | $C_d$ (GEP) | $C_d$ (NLR) |
|---|---|---|---|---|---|---|---|
| 30  | 1.082 | 0.092 | 0.011 | 0.003 | 0.86  | 0.892 | 0.847 |
| 30  | 1.082 | 0.092 | 0.017 | 0.006 | 0.76  | 0.794 | 0.709 |
| 30  | 1.082 | 0.092 | 0.026 | 0.009 | 0.684 | 0.693 | 0.611 |
| 30  | 1.082 | 0.092 | 0.032 | 0.012 | 0.625 | 0.611 | 0.534 |
| 60  | 0.56  | 0.101 | 0.013 | 0.002 | 0.872 | 0.833 | 0.803 |
| 60  | 0.56  | 0.101 | 0.031 | 0.006 | 0.705 | 0.709 | 0.684 |
| 60  | 0.56  | 0.101 | 0.051 | 0.011 | 0.573 | 0.596 | 0.588 |
| 60  | 0.56  | 0.101 | 0.029 | 0.006 | 0.713 | 0.725 | 0.701 |
| 90  | 0.396 | 0.103 | 0.014 | 0.002 | 0.789 | 0.798 | 0.762 |
| 90  | 0.396 | 0.103 | 0.047 | 0.008 | 0.702 | 0.687 | 0.685 |
| 90  | 0.396 | 0.103 | 0.069 | 0.012 | 0.572 | 0.6   | 0.607 |
| 90  | 0.396 | 0.103 | 0.058 | 0.01  | 0.626 | 0.64  | 0.639 |
| 120 | 0.323 | 0.106 | 0.027 | 0.003 | 0.791 | 0.773 | 0.744 |
| 120 | 0.323 | 0.106 | 0.044 | 0.007 | 0.74  | 0.73  | 0.710 |
| 120 | 0.323 | 0.106 | 0.073 | 0.012 | 0.665 | 0.646 | 0.648 |
| 120 | 0.323 | 0.106 | 0.06  | 0.01  | 0.697 | 0.682 | 0.672 |
| 150 | 0.29  | 0.108 | 0.014 | 0.001 | 0.797 | 0.786 | 0.785 |
| 150 | 0.29  | 0.108 | 0.071 | 0.011 | 0.698 | 0.682 | 0.662 |
| 150 | 0.29  | 0.108 | 0.034 | 0.004 | 0.796 | 0.766 | 0.731 |
| 150 | 0.29  | 0.108 | 0.052 | 0.008 | 0.736 | 0.728 | 0.694 |
| 180 | 0.28  | 0.1   | 0.055 | 0.007 | 0.656 | 0.685 | 0.653 |
| 180 | 0.28  | 0.1   | 0.072 | 0.011 | 0.675 | 0.664 | 0.643 |
| 180 | 0.28  | 0.1   | 0.045 | 0.005 | 0.66  | 0.693 | 0.666 |
| 180 | 0.28  | 0.1   | 0.061 | 0.008 | 0.68  | 0.68  | 0.652 |

Considering the estimation of coefficient of discharge relation and discharge equation on sharp-crested weir under free flow in channel, defined as follow, equation (7) shows the outflow as:

$$Q = \frac{2}{3} C_d \sqrt{2g} L y^{1.5} \quad (14)$$

where $C_d$ is coefficient of discharge, w/y the ratio of crest height to head over the crest of the weir, L/W ratio of crest length of water to channel width, L/y the ratio of crest length of water to the head over the crest of the weir, F Froude number, L crest length of water, y head over the crest of the weir, g acceleration due to gravity and $\theta$ vertex angle.

## 6      Conclusions

There are many ways to control flood such as using weirs which are either located aside or along the channel. To predict the coefficient of discharge of a weir along the channel, the present study made use of the ratio of crest height to head over the crest of the weir (p/y), crest length of water to channel width (L/W), crest length of water to the head over the crest of the weir (L/y), Froude number (F=V/√(gy)) and vortex angle (θ) and an equation has been presented as equation 7 using GEP. The accuracy of the presented model was examined through taking different statistical indexes into consideration and the results indicated that equation 7 predicts discharge coefficient with an approximate relative error of 2.5% for hydraulic conditions which had no role in training the model. Also, the amounts of all the $C_d$ predicted through using this method had a relative error less than 5%. Following that, different models were presented in order to examine the effect of each of the dimensionless parameters presented in this study. The results demonstrate that vortex angle (θ) parameter had lesser effect in predicting $C_d$ in comparison with the other models. Also, the simultaneous use of crest height ratio to the head over the crest of the weir (p/y), crest length of water to channel width (L/W), crest length of water to head over the crest of weir (L/W), Froude number (F=V/√(gy)), and vortex angle (θ) dimensionless parameters is necessary in predicting the discharge coefficient. Then, in order to examine the accuracy of the models presented by using GEP, in comparison with nonlinear regression analysis (NLR), an equation was presented through using NLR as equation 13 and the results indicated the higher accuracy of GEP in comparison with NLR.